\documentclass[12pt,preprint]{aastex}

\slugcomment{Submitted to the \apj }

\shorttitle{Testing Gravity in the Outer Solar System}
\shortauthors{Wallin et al.}

\begin{document}


\title{Testing Gravity in the Outer Solar System:
Results from  Trans-Neptunian Objects}


\author{John F. Wallin\altaffilmark{1}, David S. Dixon\altaffilmark{2}, \and Gary L. Page\altaffilmark{3}  }

\altaffiltext{1}{George Mason University, College of Science, Department of Computational and Data Sciences, Department of Physics and Astronomy, and Center for Earth Observing and Space Research (CEOSR), 4400 University Drive, MS 6A2, Fairfax, VA 22030; jwallin@gmu.edu.}
\altaffiltext{2}{Jornada Observatory, Las Cruces, NM; ddixon@cybermesa.com.}
\altaffiltext{3}{George Mason University, College of Science, Department of Computational and Data Sciences, 4400 University Drive, MS 6A2, Fairfax, VA 22030; gpage@gmu.edu.}



\begin{abstract}
The inverse square law of gravity is poorly probed by experimental
tests at distances of $\sim$ 10 AUs.   
Recent analysis of the trajectory of the Pioneer 10 and 11
spacecraft have shown an unmodeled acceleration directed
toward the Sun which was not explained by
any obvious spacecraft systematics, and occurred when at
distances greater than 20 AUs from the Sun.
If this acceleration represents a departure from Newtonian gravity
or is indicative of an additional mass distribution in the outer solar system,
it should be detectable in the orbits of Trans-Neptunian Objects (TNOs).
To place limits on deviations from Newtonian gravity,  we have selected a
well observed sample of TNOs
found orbiting between 20 and 100 AU from the Sun.
By examining their orbits with modified orbital fitting
software, we place tight limits on the perturbations
of gravity that could exist in this region of the 
solar system.

\end{abstract}





\section{Introduction}

The theory of General Relativity (GR) has been
verified with a wide variety highly sensitive 
of experiments.  The effects of time dilation,
gravitational radiation (via timing of binary pulsars), 
and gravitational lensing have been tested to very high precision.   
However, most of the experiments that test GR are in the strong
limit of gravity, where the gravitational field and
associated mass density are typical for stars and compact
objects.   Even in the solar system, we see the effects
of GR on the precession of Mercury's orbit as well as in other 
precision experiments.   However, in the weak limit of 
gravity when objects are moving slowly, GR reduces to the familiar
Newtonian form of the inverse square law  (\cite{2006LRR.....9....3W}).
This law is used in orbital dynamics to predict the location of planets
with objects more than $\sim$ 1 AU from the Sun without including
relativistic corrections.
Although Newtonian gravity's inverse square law shows excellent
agreement with observed data throughout on scales of a few AUs,
testing gravity in the outer 
solar system at distances greater than 20 AUs has been difficult.

Since objects (TNOs in particular) in the outer
part of the solar system cannot be observed with radar,
determining their orbits is done using optical astrometric
observations coupled with limited spacecraft observations.   
The accuracy of these observations
and the relatively long time span needed to observe the
outer planets has led to some difficulties in matching
their orbits to Newtonian gravity.  Even after the discovery
of Pluto, the anomalies in
Neptune's orbit were attributed to
a perturbing 10th planet, until this
issue was resolved with modern measurements
of planetary mass obtained from spacecraft (\cite{talmadge88}).
The limited astrometric accuracy, the long orbital period, and
relatively short time since most of these objects have been
accurately observed has led to uncertainty their orbits and precluded using
them for accurate tests of the inverse square law.

Other tests of the weak limit of gravity at distances greater than
$\sim$ 10 AU have generally met with limited success.  The flat
rotation curves of galaxies, for example, have been generally
interpreted as evidence of dark matter.  However, we have not yet
directly detected dark matter particles by any observational or
experimental technique.  This has led some to interpret the flat
rotation curves of galaxies as possible evidence that the Newtonian
approximation breaks down in the weak field limit.  Instead of
invoking the existence of dark matter, Modified Newtonian Dynamics
(MOND) (\cite{milgrom83}) was developed to provide an alternative
explanation of the observed flat rotations curves.  This theory has
had good success at modeling the rotation curves of many galaxies
based only on the distribution of the old stellar population.
\cite{TeVeS} has presented a Lorentz-covariant theory of gravity known
as TeVeS that yields MOND in its weak field limit.  Although MOND, TeVeS and
other alternative theories of gravity have not been verified, the idea
of Newtonian gravity breaking down in its weak limit must be
considered as an alternative to dark matter to explain galaxy rotation
curves.

The orbits of periodic comets in our solar system also have
shown deviations from Newtonian gravity.  These deviations
 have been characterized as non-gravitational
forces (\cite{kro04, mar69, mar73, mil99}) and are generally 
attributed to out-gassing
of the comets as they approach the Sun.  Each comet that shows
these deviations from Newtonian motion are fit to a set of
three acceleration parameters based on astrometric observations 
of the orbit, not on physical models of cometary out-gassing.
Because of the parametric nature of the fitting process, 
the orbits of long period comets cannot
confirm that Newtonian gravity is consistent with orbits in the
outer part of the solar system.

One obvious way to measure gravity in the outer solar system is by
using the high accuracy tracking data of spacecraft leaving the solar
system.  When the Pioneer 10 and 11 spacecraft were about 20 AU from
the Sun, their tracking data showed a systematic unmodeled
acceleration of $(8.74\pm 1.33) \times 10^{-8} $ cm s$^{-2}$ directed
toward the Sun.  This acceleration appears at 
between 10 and 20 AU, and 
then remains constant outside of about 20AU.   
The analysis of this data is detailed in
\cite{and98,and02a,and02b}.  Obvious explanations such as interactions
with the solar wind, scattering of diffuse gas off a warm spacecraft,
and electromagnetic effects have been considered.  Thus far, there are
no convincing physical phenomena that can cause this acceleration.
Although unmodeled spacecraft systematics are the most likely
explanation, it is possible that some new physical phenomenon may be
responsible for this effect.  There are currently preliminary plans
to develop a spacecraft to investigate the Pioneer Anomaly
directly (\cite{PA-Collaboration}).  Additionally, there is on-going
work to reanalyze the Pioneer spacecraft tracking data  (c.f. \cite{Turyshev2006a},
\cite{Turyshev2006b}, \cite{Turyshev2005a}) and there is
considerable debate about the meaning of the original results.  Even so, it is
certainly clear that separating the effects of spacecraft dynamics
from gravitational deceleration is a difficult task when one is trying
to measure small deviations from Newtonian gravity.

\cite{2003Icar..165..219W} have looked at the 
orbits of Oort Cloud comets in order to independently examine the Pioneer 
effect.  If the Pioneer effect was affecting comets, the gravitational
binding energy would be higher and galactic tides could not 
play the dominate role in making these objects observable.  

In this paper, we use Trans-Neptunian Objects (TNOs) to place limits
on deviations from Newtonian gravity in the outer parts of the solar system.
The use of planetary orbits to  measure these
deviations from Keplerian orbits is not new and similar analyses
have been completed by other authors using astrometric data on the
major planets
(c.f. \cite{talmadge88}, \cite{hogg91}, \cite{sereno06}, \cite{iorio06})
In these papers, the authors either examine the residuals of
the orbital fit or project the orbital trajectories
forward in time and look for the expected deviation
between existing theories and observational uncertainty.
There are limitations on both of these approaches.

In the work by Iorio (\cite{iorio06}), the authors project orbital
paths from a set of orbital elements forward with and without the
Pioneer effect and show that the differences in position are well
outside the astrometric uncertainty.  As we have shown in our
previous paper (\cite{2006ApJ...642..606P}), one must consider how
both the old and new observed positions will change the derived
orbital elements rather than just looking for a shift between
prediction and observation positions.  When new astrometric positions
for an orbiting object are obtained, a new fit to the orbital elements
is created.  If an external force perturbs an orbit, the values of the
elements will slowly change as the fitting algorithm attempts to
integrate the new data into the orbital model.  If these elements
change slowly enough, the addition of new astrometric data may produce
values that are within the uncertainty of the original values of the
orbital elements.  Of course, the newly fitted orbital parameters will
provide a good approximation to the perturbed orbit, and the
perturbations will go undetected unless the residuals are examined
over the entire orbital history.  Just projecting the resulting from a
set of static orbital elements forward in time does not address the
more subtle issues of orbital dynamics which could mask the detection
of orbital perturbations that we are attempting to detect.

Although anomalous
accelerations of a sufficient magnitude would certainly
show up in the residuals of the orbital fit, 
it is difficult to directly relate systematic 
variation in the residual to an upper limit on any 
perturbative acceleration.  
As far as we know, no one has conducted the analysis of the uncertainty
in any perturbing acceleration added to Newtonian gravity
directly from planetary observations and related them
back directly to the expected systematic change in the 
residuals of the orbital fit.  
The ``detection by modeling''
method (\cite{hogg91}) is general more sensitive than looking for
systematic changes in the residuals.   However, the very use
of a specific model for orbital perturbation
can limit the types of residuals that 
are being detected.  Thus far, most of the searches for
orbital perturbations have been looking
for a localized planet rather a radially dependent distribution
of matter or deviations from the inverse square law.

In this paper, we take advantage of the large body 
of astrometric data that has recently become available
on Trans-Neptunian Objects (TNOs).
Our approach is to use an ensemble of objects that have
been found in the outer solar system and whose observations are  
archived in the Minor Planet Center Extended Computer Service (ECS).
Extending the modeling technique
of \cite{2006ApJ...642..606P}, we fit the orbits using a modified orbital
fitting program that allows a radially directed force of arbitrary
strength to be added to gravitational accelerations already 
calculated by the program.   For each object,
we calculate this anomalous acceleration 
along with a statistically derived error
using the well documented Bootstrap technique (\cite{wall03}, \cite{efron}).

Even though most of these TNOs have only been 
recently identified, some of them have long observational arcs
because of the reanalysis of 
archival images.   The ensemble of observations used in
this paper covers a combined total
of 562 years of observations over 24 objects, making it a very
sensitive data set for examining gravity in the outer solar system.

Beyond the results of this study, the methodology presented in this
paper can be extended to new objects discovered with future large
sky surveys such as Pan-STARRS and LSST.   Using this technique,
strict limits on the deviations from Newtonian gravity can be
found constraining the solar system dark matter distribution
as well as other alternative theories of gravity such as
MOND or TeVeS.

\section{Methodology}

\subsection{The Sample}

To investigate possible gravitational perturbations to 
the inverse square law in the outer solar system,
we formed a sample of objects from the ECS.  
Our sample was selected based on three criteria:
\begin{enumerate} 
\item The object must be observed at least 20 AU from the
Sun, where the Pioneer anomaly  was detected by \cite{and02a}.
\item The object must have been observed over at least seven
oppositions at a heliocentric distance greater than 20 AU.  
\item There must be at least forty archived observations
of the object.
\end{enumerate}

The first constraint is imposed because
the Pioneer anomaly was first unambiguously detected
in the spacecraft tracking data when it was more
than 20 AU from the Sun.  The last two constraints
were derived empirically.   Our analysis has shown that
orbits with less than forty observations over at 
least seven oppositions simply are not well enough constrained
to produce accurate values of the orbital elements including
the perturbing acceleration.
When additional objects are included, the large errors associated with
their fits make them extraneous to the final weighted average of the
results.

Using the first criterion, we searched the 2006 May 1 Minor Planet Center's
ECS database of planetary orbits (MPCORB.DAT) and extracted an 
initial sample of 31 objects from the 294,488 entries.   
Observational data for each of these objects was then extracted
from the Minor Planet Center's observational archives (mpn.arc), 
and preliminary orbits
were fit using the OrbFit Consortiums OrbFit (version 3.3) program.
 Using these fitted orbits, we rejected an additional
seven objects as unsuitable for our analysis
because they failed the second and third criteria.  
These resulting list of twenty-four objects and their orbital
characteristics are listed in Table \ref{tbl-1}.

The model we are fitting to these data is very simple, and is
applied separately to each object in our sample.  We use the
bootstrap technique to estimate errors in our
fits, and then explore the results for systematic trends based
on position and orbital parameters.   Finally, we combine the
results to place a limit on deviations from the inverse square
law in the outer solar system using the ensemble of data.

\subsection{Orbital Fitting}

To search for perturbations on Newtonian gravity, we used a modified
version of the OrbFit program that is used to fit orbits to
observations of asteroids.  This code is well documented and is 
widely used in the field.  For our study, we add an
additional term to the gravitational acceleration from the Sun.  The
effective acceleration of gravity from the Sun becomes:
\begin{eqnarray} \label{eqn-1}
g_{eff} = -\left\{ \frac{G M_\odot}{R^2} + \kappa \right\} \hat{r}
\end{eqnarray}

where 
\begin{eqnarray} \label{eqn-2}
\kappa = \left\{
\begin{array}{cc}
0, &{\rm\ \ } R< 20 {\rm \ AU} \\
\epsilon, &{\rm\ \ } R \ge 20 {\rm \ AU} \\
\end{array}
\right.
\end{eqnarray}

Where $\epsilon$ is an arbitrary parameter we fit to the observed data
for each object. 

 Although we realize this model is not physically 
realistic, we adopt is based on four considerations.  First of all, the
model is consistent with what was seen in the Pioneer data and
other solar system constrains on the inverse square law.  The
anomalous acceleration is constant after approximately 20 AU.  The
Pioneer tracking data shows this anomaly turns on between 10 and 20 AU
from the Sun (c.f. \cite{and04}, Fig. 7). The particular form of the
transition is not well constrained by data.  We also know that
the inverse square law is well characterized in the inner solar system,
and more poorly constrained in the outer solar system.  The use
of a transition fits this behavior.  
Second, the model is very simple with only one
free parameter.  Since we fit this equation to each object separately
and then later examine its dependence on a set of orbital parameters,
we are making very minimal assumptions about the any anomalous
perturbation.  Since most objects are found within a narrow range of
distances from the Sun, fitting each object separately allows us to
investigate an possible dependence of $\epsilon$ on heliocentric
distance.  This is also true with variables such as ecliptic longitude
and orbital parameters such as eccentricity.  
Third, only one object
in our sample ever goes inside the 20 AU cutoff.  The inner and outer
orbital radii during the observational arc of each object in our
sample is presented in Table 1.  Including a
more complicated transition would add unnecessary complexity to the
fit and add no significant knowledge to gravitational perturbation in
the outer solar system.  Finally, the single
object (42355) that does go inside 20 AU has a large error on the
final fit of $\epsilon$, and does not significantly bias our final
results.  Thus, the final fit we are using is effectively
gravitational acceleration plus a fitted radially directed
acceleration.  The representation of the transition region at 20 AU
has no significant effect in our conclusions.

Because we have introduced the new parameter $\epsilon$ into the code,
we converge on the best value of this perturbing acceleration by using a
modified bisection method to find value of $\epsilon$ that produces
the minimal residual.  Although
there may be exceptions, brute force examination of the residual on
selected test cases has shown that there is a single value for this
minimum residual, and the values smoothly decrease toward this
minimum.  Using this bisection method, 
we are able to converge to a value
for the perturbing force at a suitable accuracy with only about twenty
iterations.  The best fit is estimated by fitting a parabola to
the three points nearest the minimum
and interpolating the location of the minimum.   
For any given set of observations, we calculate the value of $\epsilon$
along with the minimal residual.

It is important to note that adding an extra parameter to any
model will inevitably lead to non-zero values in that parameter
in poorly characterized data.   As we discuss in the next section,
it is critical to be able to
characterize the quality of the data before making conclusions
about the overall value of the parameter $\epsilon$.

\subsection{Statistical Analysis and Reliability of the Results}
In order to have confidence in the results from this study,  
we use the Bootstrap method to re-sample the observational files.
As described in \cite{efron}, there are two basic versions this
technique that can be applied to fits of data.   

The first method of `bootstrapping the observations'
directly resamples the observational file.   
For a given observational file with $n$ entries, the bootstrap file samples the 
table of observations $n$ times with replacement.   The resulting data table
is of the same length as the original, but some entries
have been duplicated and others have been dropped.  This method has limited utility
with orbital fitting, since the resampling can fundamentally change the character
of the fit.  If, for example, the resampling drops a single critical observation from
50 years ago, the overall quality and reliability of the fit will be substantially diminished.   

 The second method of `bootstrapping the residuals'
 initially fits the orbit using all the original observational
entries creating a model orbit.   The residuals of the fit
are then resampled and added to the model orbit, creating 
a set of synthetic observations.   These synthetic observations
are created at the same time intervals as the original data.
For orbital data, this method is preferable since it doesn't 
introduce the systematic bias that would occur from dropping and
duplicating observations.

In both methods, 
a new orbit is then fit to the synthetic observations.   
The process is repeated, and the acceleration parameter $\epsilon$
is tabulated.  A mean and standard deviation for $\epsilon$
is then derived from the ensemble of runs.
Details of this method and its statistical basis are discussed 
elsewhere (\cite{efron}, \cite{wall03}).   As Wall and Jenkins observe,
this technique, which seems to give something for nothing,
is well established.   Additionally, the bootstrap method has been shown
to provide converging estimates to the underlying
statistical properties of the resampled data.

For our analysis using the bootstrap methods, 
we created a set of 100 simulated orbits for each object.   
When we bootstrapped the residuals of these orbits (method two
from above), the runs all converged and gave us an estimate of the
anomalous acceleration $\epsilon$.   

Bootstrapping the observations (method one from above) was 
more problematic.
For some objects, some of the synthetic orbits
failed to converge because of the nature of the resampled
observations.   In some cases, entire years of observational measurements
can be dropped because of the resampling being done in the method.  At the
same time, duplicate observations are created giving extra weight
to arbitrary entries.  The resampling inherent with
directly bootstrapping the observations can lead to 
large gaps in the observational
arc that make the trial data sets fall below the criteria
of seven oppositions with forty observations we set for
our sample selection.   Although we do not
reject these runs a priori, the results can be a failure
to find a robust orbit that fully converges    .
Nevertheless, about 75\% of the runs that bootstrapped
the observations did converge in our analysis.

Although we believe this lack of convergence in some of our runs will not 
likely lead to a significant bias in our results, we present
the results from bootstrapping the observations only for completeness.
As discussed above, the objects that have the highest fraction
of non-converging runs were those that have the shortest
and poorest sample of observational arcs.   It is likely that
the values of the anomalous acceleration ($\epsilon$) and the 
errors ($\delta \epsilon$) are being under-estimated
on these objects.  
All eight of the objects that had convergence
rates of less than 90\%  has error estimates of greater
than 100 times the Pioneer effect.
Since the best sampled orbits have errors so much smaller than those
that only marginally  fit our criteria, the impact on our
final results is small.   We further discuss the impact of the non-converging
runs in Section \ref{results}.

It is important to reiterate that the convergence problems were not present when 
we bootstrapped the residuals.  Since the time intervals and data were
much more consistent with the original data fits, the fitting process
was much more robust.   For parametric fits, bootstrapping the residuals
is generally preferred over bootstrapping the observations
because of these issues (\cite{efron}).   Although we present the results
from both methods, we believe the results from the bootstrapped residuals
are more reliable.



\section{Results and Analysis \label{results}}

The results of the analysis using the bootstrapped residuals method
are shown in Table  \ref{tbl-2}.
The second column in this table represents
the average acceleration ($\epsilon$ in equation \ref{eqn-1})  and
its standard deviation from the 100 run ensemble.
We also calculate the average residual and standard deviation
of the residual for the ensemble of runs.  
In the `acceleration' column, we present the fitted anomalous 
acceleration using all the data in the observational files 
without using bootstrapping.  We refer to this fit as our ``best fit''
model.   No formal error can be found on for this ``best fit''
model, since they are derived from a single set of observations.
The last column represents the residual
we found by fixing $\epsilon=0$, thus providing no perturbative force.

All the forces in the table and the text below
are measured in terms of the Pioneer
effect acceleration of $8.74 \times 10^{-8}$ cm s$^{-2}$.   An acceleration of 
one in these units would be expected if the Pioneer Anomaly was affecting the 
orbits of these objects, while an acceleration of zero would 
indicated consistency with standard Newtonian gravity.

\begin{figure} 
\epsscale{1.0}
\plotone{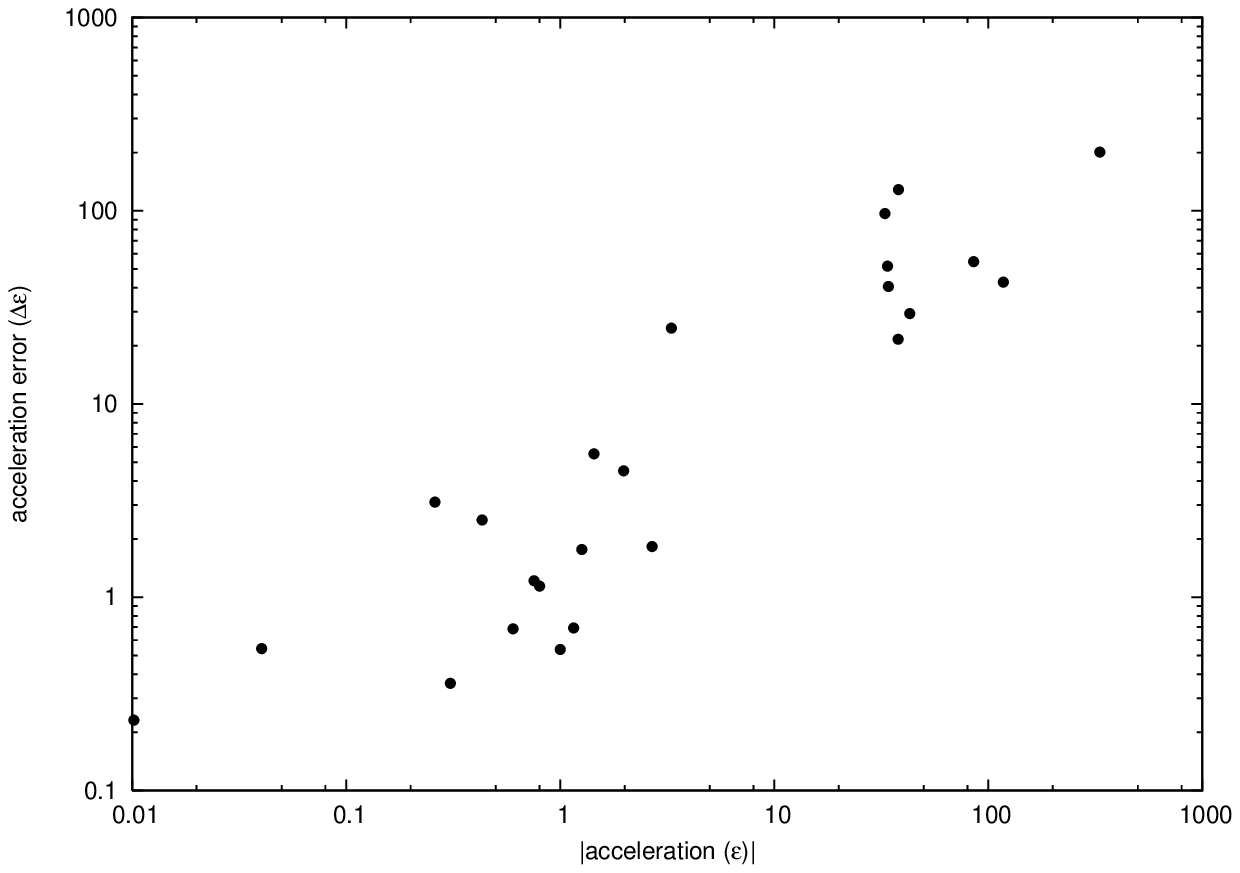}
\caption{ Error of the measured anomalous acceleration $\Delta \epsilon$ vs the magnitude of measured 
anomalous acceleration $|\epsilon|$.
The error of the measured acceleration is correlated with the strength of the acceleration, indicating
the true value of the acceleration is small.   All accelerations are measured
in units of the measured Pioneer Anomaly ($8.74 \times 10^{-8} $ cm s$^{-2}$).
Errors were derived using one standard deviation from one hundred
trial runs of the bootstrap analysis for each object.
\label{fig-1}
}
\end{figure}

To examine the consistency of our results, in Figure \ref{fig-1}
we plot the measured error in the anomalous acceleration ($\delta \epsilon$)
against the absolute value of the anomalous acceleration 
($\epsilon$) for each
object in our sample.   In this plot,
there is an obvious correlation between the error and the value of the
anomalous acceleration derive from our fits.  
The objects with large accelerations are those with 
large errors in our sample, suggesting the true value of the 
acceleration $\epsilon$ should be near zero.  

\begin{figure} 
\epsscale{1.0}
\plotone{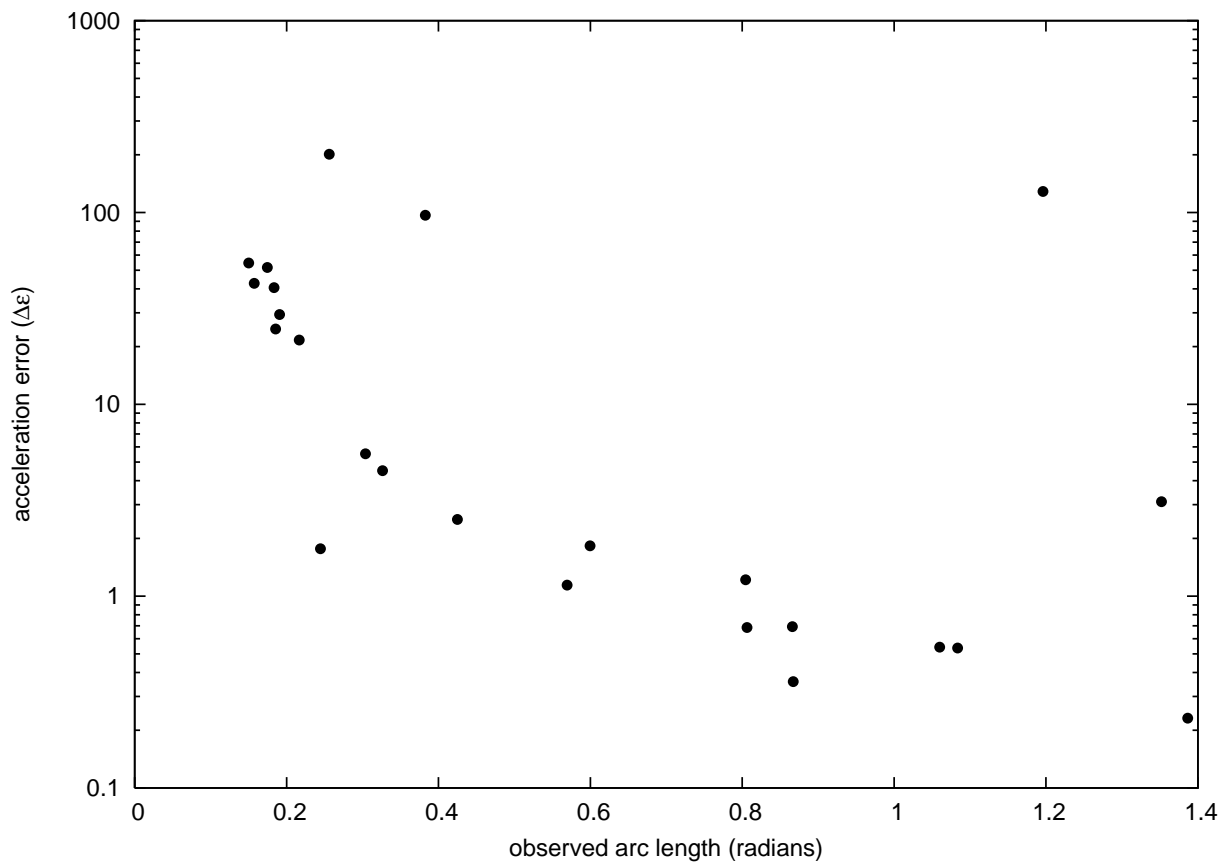}
\caption{ Error of the measured anomalous acceleration ($\Delta
\epsilon$) vs the observed geocentric arc of observations measured in
radians.  Objects with very short observational arcs ($< 0.2$) radians
are poorly characterized and do not have errors small enough to
contribute to the analysis.  A long arc-length of observations by
itself is not sufficient for small errors in the orbital fit.
Additional factors such as the sampling rate and the quality of the
astrometry play a role the reliability of the orbital fit as well.
All accelerations are plotted in units of the measured Pioneer Anomaly
($8.74\times 10^{-8} $ cm s$^{-2}$).  The acceleration errors $\Delta
\epsilon$ were derived using one standard deviation from one hundred
runs of the bootstrap analysis for each object.
\label{fig-2}}
\end{figure}

Figure \ref{fig-2} shows the relationship between the measured error
in the anomalous acceleration and the observed observational
arc in radians.   As expected, the best characterized objects
have longer observational arcs.   However, other factors such as
frequency of sampling also play a critical role in reducing the
errors in orbital determination.   Regular observations over
a long time period are likely to yield significant improvements
of these results.

In both Figure \ref{fig-1} and Figure \ref{fig-2}, the results are what we 
would have expected in this experiment.   Specifically, the estimates 
we make of our errors seem to be consistent with the behavior of the 
estimates of the accelerations.   Based on the results, we believe error
estimates we derive using the bootstrap technique provides us with 
a reliable measurement of the error associated with the fitted values
of $\epsilon$ on each 
object.    With an estimate of the
errors on $\epsilon$, we can calculate the weighted average acceleration 
using the inverse of the variance as the weights (\cite{bevington}).   
By doing this, we are assuming fitting
$\epsilon$ to a single value for all the objects.   We will examine the
validity of this assumption below.

The weighted average for the ensemble of bootstrapped residual runs was 
$0.10 \pm 0.16$   times the Pioneer acceleration where the
uncertainty is a one-sigma error.  If we use
the bootstrap errors for the weighting and use the ``best fit''
models as the values for the run, the average acceleration
is identical to two significant digits.   
These results are consistent with standard Newtonian gravity and a value of zero
for $\epsilon$.   Using the `bootstrapped observation' method, we find the ensemble of bootstrap runs
gives us an average acceleration of  $-0.23 \pm 0.28$ times the Pioneer acceleration.
If we use
the bootstrap errors for the weighting and use the ``best fit''
 models as the values for the run, the average acceleration
 is $0.03 \pm 0.28$.     When we only include the objects that have had no problems with
convergence in any of their runs, the final results we obtain
for the acceleration is identical to three significant digits
as the weighted average of the overall sample.   The lack
of convergence in the some runs using the bootstrapped observation method
creates no significant bias in our results.

\begin{figure} 
\epsscale{1.0}
\plotone{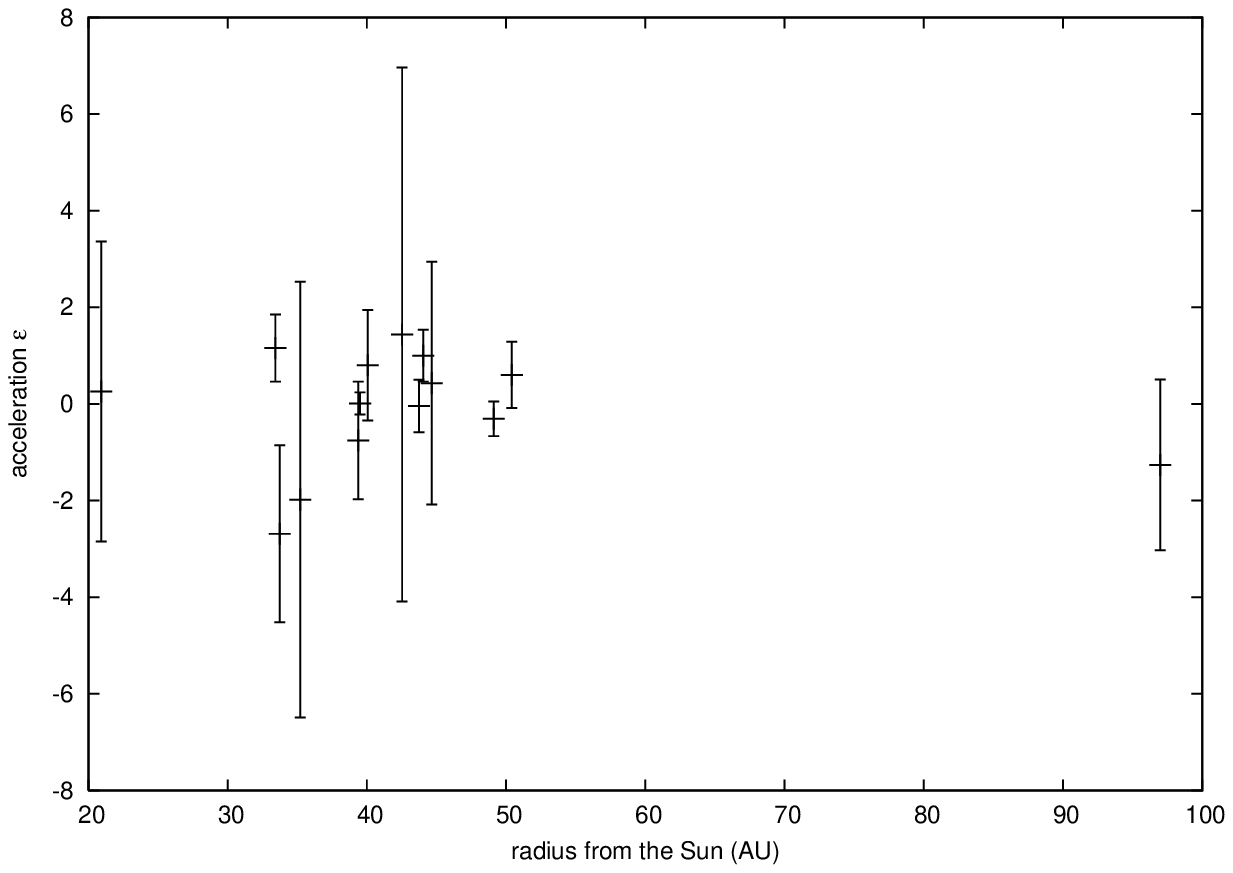}
\caption{ Anomalous acceleration $\epsilon$ vs average distance of the object from the Sun in AU over
the observed arc.
There is no apparent trend between the measured anomalous acceleration and the distance from the
Sun.  Only the objects with errors less than $\pm 10$ times the acceleration of the 
Pioneer effect were included in this plot for clarity.  All accelerations are plotted
in terms of the measured Pioneer Anomaly ($8.74\times 10^{-8} $ cm s$^{-2}$).
The acceleration errors ($\Delta \epsilon$) were derived using one standard deviation from one hundred
trial runs of the bootstrap analysis for each object.
\label{fig-3}
}
\end{figure}

To further examine gravity in the outer solar system, we looked at how
the measured acceleration is correlated with the position and orbital
parameters of the objects in our sample. Figure \ref{fig-3} shows the
relationship between the measured anomalous acceleration
($\epsilon$) and the derived heliocentric distance of the orbits of
our objects.  For clarity, only the objects with accelerations errors
less than ten times the Pioneer Anomaly were included in the plot.
There is no clear correlation between the heliocentric distance and
the anomalous acceleration.  However, the error bars on our
measurements are too large to completely rule out such a correlation.
  
\begin{figure} 
\epsscale{1.0}
\plotone{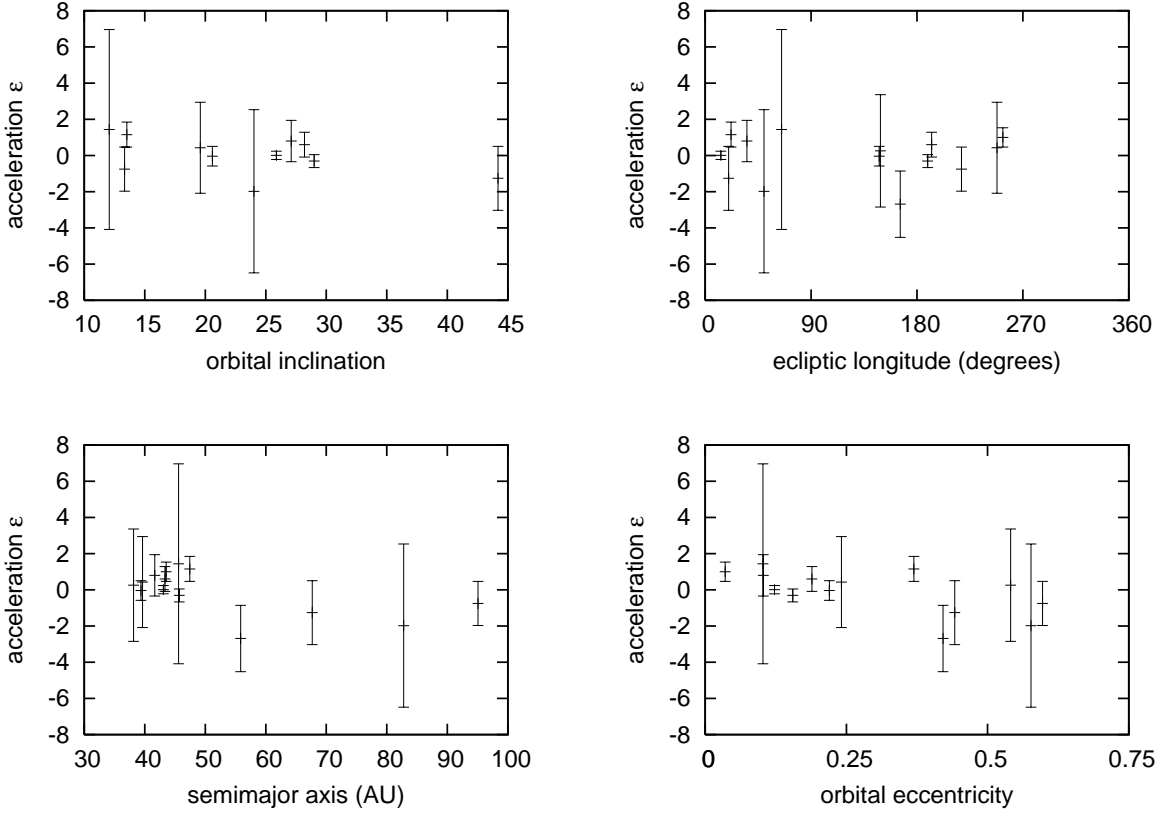}
\caption{ The measured anomalous acceleration ($\epsilon$) vs orbital inclination, 
 ecliptic longitude, semimajor axis, and orbital eccentricity.   
 Only the objects with errors less than $\pm 10$ times the acceleration of the 
Pioneer effect were included in this plot for clarity.
There is no evidence of any statistically significant
correlation between these four parameters and
the corresponding acceleration.  
 All accelerations are plotted 
in terms of the measured Pioneer Anomaly ($8.74\times 10^{-8} $ cm s$^{-2}$).
The acceleration errors ($\Delta \epsilon$)  were derived using one standard deviation from one hundred
trial runs of the Bootstrap analysis for each object.
\label{fig-4}}
\end{figure}

In Figure \ref{fig-4}, we compare the anomalous accelerations for these TNOs 
with the eccentricity of the orbit,
semi-major axis of the orbit, average heliocentric distance
during the observations, and ecliptic longitude.
No statistically significant correlations
were found in the data between mean acceleration $\epsilon$ and any of these
parameters.   There is a weak correlation seen between semimajor axis and
$\epsilon$ as well as orbital
eccentricity and $\epsilon$.   Although this should be examined 
further with future observations,
it is well within the noise and not considered significant.
Because of the large uncertainty in some of these
measurements, detecting small trends in the data is not possible.
With that caveat, there is no evidence to support any systematic deviations
from the inverse square law as a function of these variables.
In general, we find no correlation between the orbital elements 
or positions of our objects and the
results we find for the anomalous acceleration.

\section{Conclusions}

In this paper, we have presented a new method using orbital
measurements of an ensemble of TNOs to measure deviations
from the inverse square law or gravity in the outer solar system.  The method
relies on doing separate orbital fits for each object, and then
characterizing the accuracy of each fit using the bootstrap technique.
Since no significant systematic trends were detected in our
sample, we combined
the data from all the objects using a weighted average to place
limits on deviations from the gravitational inverse square law
in the outer solar system.
Using existing
data, we have measured the deviation from the inverse square law
to be  $\delta a = 8.7 \times 10^{-9} \pm 1.6 \times 10^{-8}$
cm s$^{-2}$ directed outward from the Sun for objects at
heliocentric distances of 20 to 100 AUs.
This result is consistent with zero at the 1$\sigma$ limit.

Based on our analysis of the observational data of TNOs, 
we find that the gravitational acceleration in the outer solar system
is inconsistent with the Pioneer anomaly
at the $\sim 5\sigma$ level using both variations of the bootstrap analysis.
All of our results 
were consistent (within $1 \sigma$) with Newtonian gravity
without any additional radial perturbative forces.   This suggests that the
deceleration seen in the Pioneer tracking data was probably the result
of spacecraft systematics rather than exotic physics.  
Even so, we cannot rule out the possibility that exotic physics is 
affecting the Pioneer spacecraft trajectories.   Our work only shows
that the trajectory data from the Pioneer Spacecraft is inconsistent 
to what we see in large, slowly moving rocks in the outer solar system
(\cite{2006ApJ...642..606P}).

These results were derived using existing astrometric data on minor planets 
in the outer part of the solar system.   The use of the
bootstrap method in this analysis has allowed us to provide a test
of the reliability of each orbital fits.     Although bootstrapping the observations
directly is not held to be as reliable as bootstrapping the residuals, both results
are consistent with our conclusions.   
The combination of these two methods provide a cross check, albeit a weak one, on 
our results.

We found no evidence of any correlation
between the measured values of $\epsilon$ and the object's position in the
solar system or its orbital characteristics.   However, we would have only detected such
a trend if it were strongly present in the data with an amplitude much greater
than the Pioneer effect acceleration.   

Overall, the results we present confirm the veracity of the Newtonian gravitational
potential in the outer parts of the solar system.   However, future analysis
of astrometric data from Pan-STARRS and LSST will provide a much more sensitive test
of gravity.   If
the number of TNOs in our sample is expanded by a factor of 100, and these
objects have a long arc of regularly observed positions, we will be able
to increase the sensitivity of our results by a factor of about ten, depending
on the arc lengths and rate of observations.   The data sets from Pan-STARRS
and LSST that will be created within the next ten years 
will provide strong limits on alternative gravitational theories such as MOND.

Additional future work will focus on how
well this technique works for finding accelerations using
ensembles of synthesized observations.   Of specific interest is
the role that orbital eccentricity and the length and completeness
of the observational arc plays in the results we obtain from
fitting acceleration parameters to observational data.

\acknowledgments

The authors wish to acknowledge the Minor Planet Center for
observational data, available through their Extended Computer
Service\footnote{\url{http://cfa-www.harvard.edu/iau/services/ECS.html}}.
Additionally, the excellent software packages developed and maintained
by the OrbFit consortium\footnote{\url{http://newton.dm.unipi.it/orbfit}}
allowed orbital calculations to be performed with the requisite
precision. The OrbFit program made use of JPL's DE405 ephemeris
data\footnote{\url{http://ssd.jpl.nasa.gov/eph\_info.html}} to
describe the dynamics of the solar system.  Finally, the authors would
like to think Dr. James Gentle and Dr. Daniel Carr of George Mason University
for their helpful input regarding the statistical analysis of this data.

\clearpage




\clearpage

\clearpage

\begin{deluxetable}{cccccccc}
\tabletypesize{\scriptsize}
\tablecaption{TNO Objects Investigated \label{tbl-1}}
\tablewidth{0pt}
\tablehead{
\colhead{Name} & \colhead{\# Observations\tablenotemark{a}} & \colhead{a\tablenotemark{b}} (AU) &
\colhead{e\tablenotemark{c}} & \colhead{R1\tablenotemark{d} (AU)} 
& \colhead{R2\tablenotemark{d} (AU)} & 
\colhead{Time Span\tablenotemark{e} (yr)} &\colhead{Oppositions} 
}
\startdata
15760 & 74  & 43.7402581 & 0.0656661 & 40.8835 & 40.9108 & 7.3  & 7  \\
 15788 & 46  & 39.1457785 & 0.3174450 & 32.7020 & 30.5452 & 7  & 8  \\
 15789 & 87  & 39.3665451 & 0.1845412 & 33.8839 & 34.7827 & 6.1  & 7  \\
 15807 & 44  & 44.0377756 & 0.0630997 & 41.9973 & 42.3789 & 8  & 8  \\
 15809 & 52  & 42.4362988 & 0.2198345 & 36.0171 & 34.8523 & 7  & 7  \\
 15874 & 188  & 82.7812076 & 0.5768854 & 35.2315 & 35.1883 & 9.1  & 7  \\
 15875 & 66  & 39.2009878 & 0.3275728 & 26.4840 & 26.4669 & 7  & 7  \\
 16684 & 50  & 44.3792306 & 0.0534350 & 43.1593 & 42.8591 & 7  & 7  \\
 19521 & 106  & 45.5986282 & 0.1024126 & 43.0195 & 42.0543 & 13  & 10  \\
 20000 & 86  & 42.9541121 & 0.0515448 & 40.9771 & 43.2804 & 51.1  & 10  \\
 24835 & 113  & 41.6500352 & 0.1028177 & 41.1594 & 38.9536 & 23.2  & 10  \\
 26181 & 40  & 95.0672716 & 0.5974190 & 38.9791 & 39.7789 & 24  & 7  \\
 26308 & 48  & 47.4463316 & 0.3696155 & 30.4813 & 36.3888 & 23.2  & 8  \\
 26375 & 49  & 55.8564193 & 0.4212401 & 32.3838 & 35.1070 & 16.2  & 7  \\
 28978 & 35  & 39.6233386 & 0.2411777 & 46.5949 & 42.7291 & 22  & 7  \\
 42355 & 112  & 38.1638482 & 0.5407749 & 24.3108 & 17.5387 & 16.2  & 9  \\
 50000 & 85  & 43.5479367 & 0.0354944 & 44.7689 & 43.3330 & 51.2  & 11  \\
 55636 & 43  & 43.0880405 & 0.1226813 & 37.9468 & 41.0818 & 51.3  & 7  \\
 79360 & 111  & 43.9428179 & 0.0142468 & 43.6099 & 43.5536 & 7.9  & 8  \\
 90482 & 109  & 39.3860863 & 0.2199809 & 39.8004 & 47.6875 & 54  & 10  \\
 J93F00W & 63  & 44.0349446 & 0.0543513 & 42.1583 & 41.9245 & 8.1  & 7  \\
 K03E61L & 82  & 43.3372598 & 0.1891092 & 49.5992 & 51.2233 & 51  & 12  \\
 K03UV3B & 186  & 67.6700187 & 0.4417510 & 97.0599 & 96.8932 & 51.3  & 14  \\
 K05F09Y & 110  & 45.7088051 & 0.1550054 & 46.3297 & 51.9059 & 51.2  & 9  \\

\enddata
\tablenotetext{a}{Number of observations }
\tablenotetext{b}{Semi-major axis (AU)}
\tablenotetext{c}{Eccentricity}
\tablenotetext{d}{Radius from the Sun at the beginning (R1) and end (R2) of the observations (AU)}
\tablenotetext{e}{Time span between first and last observations (years)}
\end{deluxetable}

\clearpage


\clearpage

\begin{deluxetable}{rccccc}
\tabletypesize{\scriptsize}
\tablecaption{TNO Bootstrap Analysis \label{tbl-2}}
\tablewidth{0pt}
\tablehead{
  \colhead{Object Name} 
& \colhead{Bootstrap Acceleration\tablenotemark{a}} 
& \colhead{Bootstrap Residual\tablenotemark{b} } 
& \colhead{Acceleration\tablenotemark{c}}
& \colhead{Residual  \tablenotemark{d}} 
} 
\startdata
15760 & $ 85.4 \pm 54.5  $ & $ 0.67042  \pm  0.039546 $ & 86.8 & 0.68433 \\
15788 & $ 332 \pm 201  $ & $ 0.68436  \pm  0.062450 $ & 119 & 0.80242 \\
15789 & $ -34.1 \pm 40.5  $ & $ 0.59386  \pm  0.043949 $ & -29.4 & 0.61228 \\
15807 & $ 42.9 \pm 29.4  $ & $ 0.76216  \pm  0.057195 $ & 37.7 & 0.81021 \\
15809 & $ 37.9 \pm 21.6  $ & $ 0.46411  \pm  0.042229 $ & 38.8 & 0.48921 \\
15874 & $ -1.98 \pm 4.51  $ & $ 0.48916  \pm  0.028637 $ & -2.19 & 0.50035 \\
15875 & $ 32.9 \pm 96.8  $ & $ 0.55931  \pm  0.069292 $ & 16.8 & 0.56625 \\
16684 & $ 117 \pm 42.7  $ & $ 0.44258  \pm  0.023309 $ & 121 & 0.48592 \\
19521 & $ 1.44 \pm 5.53  $ & $ 0.41250  \pm  0.034758 $ & 0.722 & 0.42752 \\
20000 & $ 0.426 \pm 0.966  $ & $ 0.47528  \pm  0.046502 $ & 0.359 & 0.49634 \\
24835 & $ 0.801 \pm 1.14  $ & $ 0.58405  \pm  0.050270 $ & 0.642 & 0.58942 \\
26181 & $ -0.754 \pm 1.22  $ & $ 0.65125  \pm  0.061349 $ & -0.671 & 0.67256 \\
26308 & $ 1.15 \pm 0.693  $ & $ 0.47548  \pm  0.030127 $ & 1.10 & 0.50099 \\
26375 & $ -2.69 \pm 1.83  $ & $ 0.39551  \pm  0.044992 $ & -2.94 & 0.40950 \\
28978 & $ 0.432 \pm 2.51  $ & $ 0.36524  \pm  0.049641 $ & 0.485 & 0.37840 \\
42355 & $ 0.260 \pm 3.10  $ & $ 0.51636  \pm  0.062756 $ & 0.378 & 0.53024 \\
50000 & $ 1.0 \pm 0.537  $ & $ 0.52599  \pm  0.042784 $ & 1.18 & 0.56972 \\
55636 & $ 0.0102 \pm 0.231  $ & $ 0.44148  \pm  0.055523 $ & -0.0168 & 0.45739 \\
79360 & $ 3.31 \pm 24.7  $ & $ 0.53585  \pm  0.038037 $ & 6.76 & 0.54377 \\
90482 & $ -0.0403 \pm 0.542  $ & $ 0.40644  \pm  0.050609 $ & -0.0404 & 0.41601 \\
J93F00W & $ 33.8 \pm 51.7  $ & $ 0.53675  \pm  0.047901 $ & 37.5 & 0.56737 \\
K03E61L & $ 0.602 \pm 0.686  $ & $ 0.41072  \pm  0.043557 $ & 0.503 & 0.41953 \\
K03UV3B & $ -1.26 \pm 1.76  $ & $ 0.4380  \pm  0.029639 $ & -1.21 & 0.44969 \\
K05F09Y & $ -0.307 \pm 0.359  $ & $ 0.23759  \pm  0.013461 $ & -0.289 & 0.23766 \\

\enddata
\tablenotetext{a}{The average of the best fit to the anomalous  acceleration $\epsilon$  in terms of the 
Pioneer Anomaly ($8.74\times 10^{-8}$ cm s$^{-1}$)
along with the 1$\sigma$ error from the Bootstrap analysis.}
\tablenotetext{b}{Average residual (arcseconds) and its 1$\sigma$ error from the bootstrap analysis.}
\tablenotetext{c}{The best fit to the anomalous  acceleration $\epsilon$  in terms of the 
Pioneer Anomaly ($8.74\times 10^{-8}$ cm s$^{-1}$).  No formal error is available
on this measurement since it was derived from a single fit.}
\tablenotetext{d}{Residual (arcseconds) of the orbit assuming no perturbing force ($\epsilon = 0$).}
\end{deluxetable}

\clearpage


\end{document}